\documentclass[twocolumn, twocolappendix, trackchanges]{aastex63}

\usepackage{amsmath}
\usepackage{booktabs}
\usepackage{longtable}
\usepackage{rotating}
\usepackage{graphicx}
\usepackage{hyperref}
\usepackage{bm}

\newcommand{\nc}[2]{\newcommand{#1}{\ensuremath{#2}\xspace}}

\usepackage{xstring} 

\newcommand{\val}[1]{%
  \IfEqCase{#1}{%
{nstars-legacy}{719}
{nstars-new}{135}
}[\PackageError{tree}{Undefined option to tree: #1}{}]%
}%

\usepackage{xspace}

\nc{\specmatch}{ \mathbf{SpecMatch} } 
\nc{\emcee}{ \mathbf{emcee} }
\nc{\radvel}{ \mathbf{RadVel} }
\nc{\rvsearch}{ \mathbf{RVSearch} }
\nc{\thejoker}{ \mathbf{TheJoker} }

\nc{\Kp}{ \textit{Kp} }
\nc{\dAICc}{\Delta\mathrm{AICc}}
\nc{\dBIC}{\Delta\mathrm{BIC}}

\nc{\rsun}{R_{\odot}}
\nc{\msun}{M_{\odot}}
\nc{\rearth}{R_{\oplus}}
\nc{\mearth}{M_{\oplus}}
\nc{\fearth}{F_{\oplus}}
\nc{\mjup}{M_{\textrm{J}}}
\nc{\Rp}{R_P}
\nc{\Mp}{M_P}
\nc{\msini}{M \sin i}
\nc{\Rstar}{R_\star} 
\nc{\Mstar}{M_\star}

\nc{\teff}{T_{\rm eff}}
\nc{\logg}{\log{g}}
\nc{\feh}{[\mbox{Fe}/\mbox{H}]}
\nc{\vsini}{V \sin i}

\nc{\kms}{\text{km\,s}^{-1}}
\nc{\ms}{\text{m\,s}^{-1}}
\nc{\msy}{\text{m\,s}^{-1}\text{yr}^{-1}}
\nc{\gmc}{\text{g\,cm}^{-3}}

\received{May 23rd, 2022}
\revised{Sep 12th, 2023}
\accepted{Oct 2nd, 2023}
\submitjournal{ApJS}

\shortauthors{Rosenthal et al.}
\shorttitle{California Legacy Survey IV: Lonely, Poor, and Eccentric}

\begin{document}
\pagenumbering{arabic}

\title{THE CALIFORNIA LEGACY SURVEY IV. LONELY, POOR, AND ECCENTRIC: A COMPARISON BETWEEN SOLITARY AND NEIGHBORLY GAS GIANTS}

\correspondingauthor{Lee J.\ Rosenthal}
\email{lrosenth@caltech.edu}

\author[0000-0001-8391-5182]{Lee J.\ Rosenthal}
\affiliation{Cahill Center for Astronomy $\&$ Astrophysics, California Institute of Technology, Pasadena, CA 91125, USA}

\author[0000-0001-8638-0320]{Andrew W.\ Howard}
\affiliation{Cahill Center for Astronomy $\&$ Astrophysics, California Institute of Technology, Pasadena, CA 91125, USA}

\author[0000-0002-5375-4725]{Heather A.\ Knutson}
\affiliation{Division of Geological and Planetary Sciences, California Institute of Technology, Pasadena, CA 91125, USA}

\author[0000-0003-3504-5316]{Benjamin J.\ Fulton}
\affiliation{Cahill Center for Astronomy $\&$ Astrophysics, California Institute of Technology, Pasadena, CA 91125, USA}
\affiliation{IPAC-NASA Exoplanet Science Institute, Pasadena, CA 91125, USA}

\begin{abstract}

We compare systems with single giant planets to systems with multiple giant planets using a catalog of planets from a high-precision radial velocity survey of FGKM stars. Our comparison focuses on orbital properties, planet masses, and host star properties. We use hierarchical methods to model the orbital eccentricity distributions of giant singles and giant multis, and find that the distributions are distinct. The multiple giant planets typically have moderate eccentricities and their eccentricity distribution extends to $e=0.47$ (90th percentile), while the single giant planets have a pile-up of nearly circular orbits and a long tail that extends to $e=0.77$. We determine that stellar hosts of multiple giants are distinctly more metal-rich than hosts of solitary giants, with respective mean metallicities $0.228\pm0.027$ vs. $0.129\pm0.019$ dex. We measure the distinct occurrence distributions of single and multiple giants with respect to orbital separation, and find that single gas giants have a $\sim$2.3$\sigma$ significant hot ($a<0.06$) Jupiter pile-up not seen among multi giant systems. We find that the median mass ($\msini$ ) of giants in multiples is nearly double that of single giants (1.71 $\mjup$ vs. 0.92 $\mjup$ ). We find that giant planets in the same system have correlated masses, analogous to the `peas in a pod' effect seen among less massive planets.
\end{abstract}


\section{Introduction} \label{sec:intro}

Giant planets are central players in planetary systems. Their dynamics affect the presence of small inner planets \citep{Zhu18, Bryan19, Rosenthal22}, and can reveal key information about the formation and dynamical history of a system. We can also better understand the impact that giant planets have on each other by studying their eccentricity distributions, as planets with non-zero eccentricities must have reached their states either by interacting with protoplanetary disks, or by undergoing dynamical interactions with other planets or stellar companions \citep{Dawson13, Petrovich16}. While secular interactions between planets and planet-disk effects can account for giants excited to moderate eccentricities ($0.2 \leq e \leq 0.6$) \citep{Dawson18}, it is more difficult to use them as an explanation for the highest-eccentricity giants that have been discovered to date, which extend to $e > 0.9$ \citep{Blunt19}.

Prior work has shown that highly eccentric giants can reach their states due to strong gravitational scattering events, possibly with other giants \citep{Chatterjee08}. This would imply that giant multiplicity is a key factor in understanding the dynamical evolution and final states of planetary systems. Perhaps systems in which we currently only see one eccentric giant began with multiple giant planets that scattered each other, one into a bound eccentric orbit and one into engulfment or an unbound trajectory. We can explore this theory by comparing the eccentricity distributions of systems with one observed giant planet and multiple observed giant planets, as well as their semi-major axis occurrence distributions and host star properties. Recent theoretical work has provided giant planet hypotheses that are easily testable with radial velocity (RV) surveys. For instance, \cite{Jackson21} used a synthetic population to show that if an RV search for outer giant companions to warm Jupiters (10--200 days) finds that a substantial fraction of these warm giants have outer companions, then we should seriously consider a formation scenario in which secular interactions cause cold giants to migrate inward and become stable warm Jupiters.

The California Legacy Survey \citep[CLS,][]{Rosenthal21} is well suited for a comparative study of single and multiple giant planets. As a three-decade long blind RV survey, it produced a sample that is appropriate for a variety of occurrence measurements, and contains over a hundred giant ($\msini\ > 0.1 M_J$) planets, in both single and multiple configurations. In this paper, we leverage this survey to compare and contrast lonely giants and multi giant systems. In Section 2, we review the star and planet catalog of the CLS. In Section 3, we describe our methods for computing planet occurrence, hierarchically modeling eccentricity distributions, and comparing the host star properties of distinct planetary samples. In Section 4, we present our results. In Section 5, we discuss our findings and their context.

\section{Survey Review}

The California Legacy Survey \citep{Rosenthal21} is a sample of RV-observed FGKM stars and their associated planets, created in order to provide a stellar and planetary catalog for occurrence studies. We approximated a quantifiably complete survey by selecting HIRES-observed stars that were originally chosen without bias towards a higher or lower than average likelihood of hosting planets. That left us with 719 stars. We used an iterative periodogram method to search for planet candidates, and performed uniform vetting using magnetic activity indicators and instrument systematics to identify false positives.

\begin{figure*}[ht!]
\begin{center}
\includegraphics[width = \textwidth]{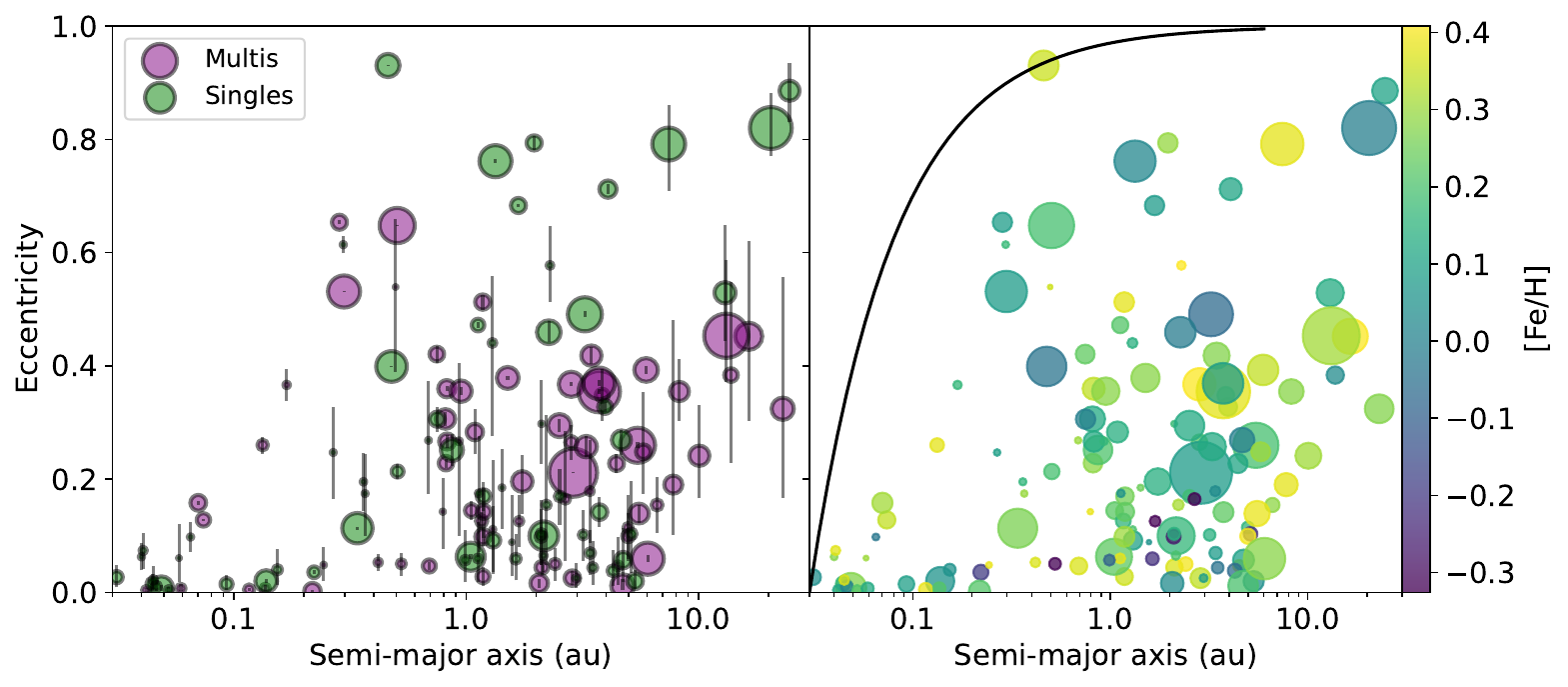}
\caption{Left: Eccentricity with respect to semi-major axis for single giant planets and multiple giant planets. Vertical bars show 15.9--84.1$\%$ confidence intervals. Circle radius is proportional to $\msini$. Right: Same as left, but color-coded by host star metallicity. Curve shows estimate of tidal disruption limit assuming that periastron at 0.03 au leads to disruption \citep{Dawson18}.}
\label{fig:axis_eccentricity}
\end{center}
\end{figure*}

\begin{figure}[ht!]
\includegraphics[width = 0.5\textwidth]{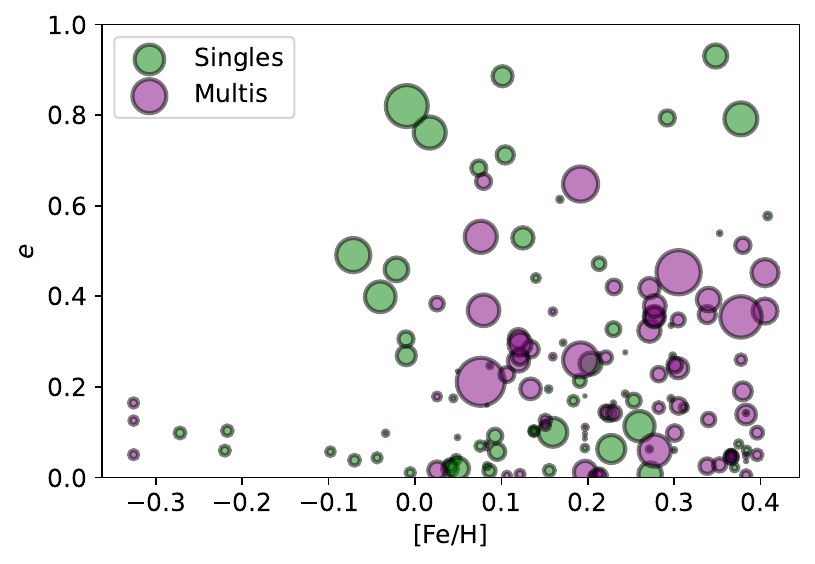}
\caption{Planet eccentricity versus host star metallicity for single and multiple giant systems.}
\label{fig:ecc_metal}
\end{figure}

This left us with 178 planets in our sample, including 134 planets with $\msini\ > 0.1 M_J$. 65 of these gas giants lack detected companions (`lonely giants'), and 69 belong to 31 multiple-giant systems. Figure \ref{fig:axis_eccentricity} shows our giant catalog in eccentricity and semi-major axis space, split between giant singles and giant multis in one panel, and color-coded by host star metallicity in another panel. Figure \ref{fig:ecc_metal} shows the catalog in eccentricity and metallicity space.

We use $0.1 M_J$ or 30 \mearth\ as our cutoff for giant planets because this is roughly the minimum mass at which runaway gas accretion can begin \citep{Lissauer09}. This is also near the threshold for planets to have compositions dominated by gaseous envelopes \citep{Lee19, Thorngren16}. These threshold bases set a more liberal definition of giant planet than one motivated by dynamical influence on planet formation and migration, but is more relevant to a planet's growth history and captures sub-Saturn planets that can still have an impact on smaller planets. Furthermore, our sample is reasonably insensitive to our definition of this threshold. When we lower the threshold from 30 \mearth\ to 20 \mearth\, the number of single giant systems only decreases from 65 to 64, and the number of multi giant systems only increases from 31 to 34.

\section{Methods}
\label{sec:methody}

\subsection{Occurrence model}

This work measures the occurrence of single and multiple giant planets as a function of multiple orbital properties. Many studies of RV or transit surveys use the intuitive occurrence measurement method known as "inverse detection efficiency" \citep{Howard12, Petigura13b}. According to this procedure, one estimates occurrence in a region of parameter space by counting up the planets found in that region, with each planet weighted by the search completeness in that region. One can measure the search completeness map of a survey by injecting many synthetic signals into each dataset, and computing the fraction of signals in a given region that are recovered by the search algorithm in use. Inverse detection efficiency is actually a specific case of a Poisson likelihood method, in which one models an observed planet catalog as the product of an underlying Poisson process and empirical completeness map \citep{Foreman-Mackey14}.

In this work, as in previous papers in the CLS \citep{Fulton21, Rosenthal22}, we used the Poisson likelihood method to model the occurrence of giant planets, taking measurement uncertainty into account. Given an observed population of observed planets with orbital and $\msini$ posteriors $\{\bm{\omega}\}$, and associated survey completeness map $Q(\bm{\omega})$, and assuming that our observed planet catalog is generated by a set of independent Poisson process draws, we can evaluate a Poisson likelihood for a given occurrence model $\Gamma(\bm{\omega} | \bm{\theta})$, where $\bm{{\theta}}$ is a vector of model parameters. The observed occurrence $\hat{\Gamma}(\bm{\omega} | \bm{\theta})$ of planets in our survey can be modeled as the product of the measured survey completeness and some underlying occurrence model,

\begin{equation}
\hat{\Gamma}(\bm{\omega} | \bm{\theta}) = Q(\bm{\omega})\Gamma(\bm{\omega} | \bm{\theta}). \\
\end{equation}

The Poisson likelihood for an observed population of objects is

\begin{equation}
\mathcal{L} = e^{-\int \hat{\Gamma}(\bm{\omega} | \bm{\theta}) \,d\bm{\omega}} \prod_{k=1}^{K} \hat{\Gamma}(\bm{\omega}_k | \bm{\theta}),\\
\end{equation}
where $K$ is the number of observed objects, and $\bm{\omega}_k$ is the $k$th planet's orbital parameter vector. The Poisson likelihood can be understood as the product of the probability of detecting an observed set of objects (the product term in Equation 2), and the probability of observing no additional objects in the considered parameter space (the integral over parameter space). Another way to understand this model is that it represents a Poisson process, with some probability of generating planets with a given mass and semi-major axis rate density.

Equations 1 and 2 serve as the foundation for our occurrence model, but do not take into account uncertainty in our measurements of planetary orbits and minimum masses. In order to do this, we use Markov Chain Monte Carlo methods to empirically sample the orbital posteriors of each system \citep{Rosenthal21}. We can hierarchically model the orbital posteriors of each planet in our catalog by summing our occurrence model over many posterior samples for each planet. The hierarchical Poisson likelihood is therefore approximated as

\begin{equation}
\mathcal{L}\approx e^{-\int \hat{\Gamma}(\bm{\omega} | \bm{\theta}) \,d\bm{\omega}} \prod_{k=1}^{K} \frac{1}{N_k} \sum_{n=1}^{N_k} \frac{\hat{\Gamma}(\bm{\omega}_k^n | \bm{\theta})}{p(\bm{\omega}_k^n | \bm{\alpha})},\\
\end{equation}
where $N_k$ is the number of posterior samples for the $k$th planet in our survey, and $\bm{\omega}_k^n$ is the $n$th sample of the $k$th planet's posterior. $p(\bm{\omega} | \bm{\alpha})$ is our prior on the individual planet posteriors. We placed uniform priors on ln($M$sin$i$) and ln($a$). We used \texttt{emcee} \citep{DFM13} to sample our hierarchical Poisson likelihood.

\subsection{Application to eccentricity}

As done in prior eccentricity studies \citep{Kipping13, VanEylen19, Bowler20}, we model the eccentricity distribution of a population of exoplanets with the Beta distribution, because it is [0, 1] bound, flexible in its shape, and has only two model parameters. The Beta distribution follows the probability density function

\begin{equation}
p(x|\alpha,\beta)\ =\ \frac{\Gamma(\alpha + \beta)}{\Gamma(\alpha)\Gamma(\beta)}x^{\alpha - 1}(1 - x)^{\beta - 1} \\
\end{equation}
where $\Gamma(x)$ is the Gamma function, rather than the occurrence rate in Equation 3. We only consider planets with \msini\ $\geq$ 0.1 \mjup\ . In other words, we are only considering the eccentricity distributions of planets with masses greater than a third of Saturn's mass.

When we apply our hierarchical modeling framework to the eccentricity posteriors of our giant planets, we produce the likelihood

\begin{equation}
\mathcal{L}(\bm{e} | \bm{\theta})\propto \prod_{k=1}^{K} \frac{1}{N} \sum_{n=1}^{N}p(e_k ^n |\bm{\theta}), \\
\end{equation}

\noindent where $\theta$ is the set of model parameters, $K$ is the number of planets, $N$ is the number of samples drawn from each planet's posterior, and $e_k^n$ is the eccentricity of the $n$th sample drawn from the $k$th planet's posterior. The normalization term in Equation (3) disappears because Equation (4) is a normalized probability density function.

\subsection{Empirical distribution comparisons}

In order to compare the host-star metallicities and other properties of distinct groups of planetary systems, we use simple statistical tests to compare the cumulative distribution functions, otherwise known as empirical distributions, of these groups. Specifically, we use the Anderson-Darling test, a statistical method for computing the probability that two empirical distributions were drawn from the same underlying distribution.

\section{Results}

\subsection{Single and multiple giants have distinct eccentricity distributions}

\label{subsec:ecc_compare}

\begin{figure}[ht!]
\includegraphics[width = 0.5\textwidth]{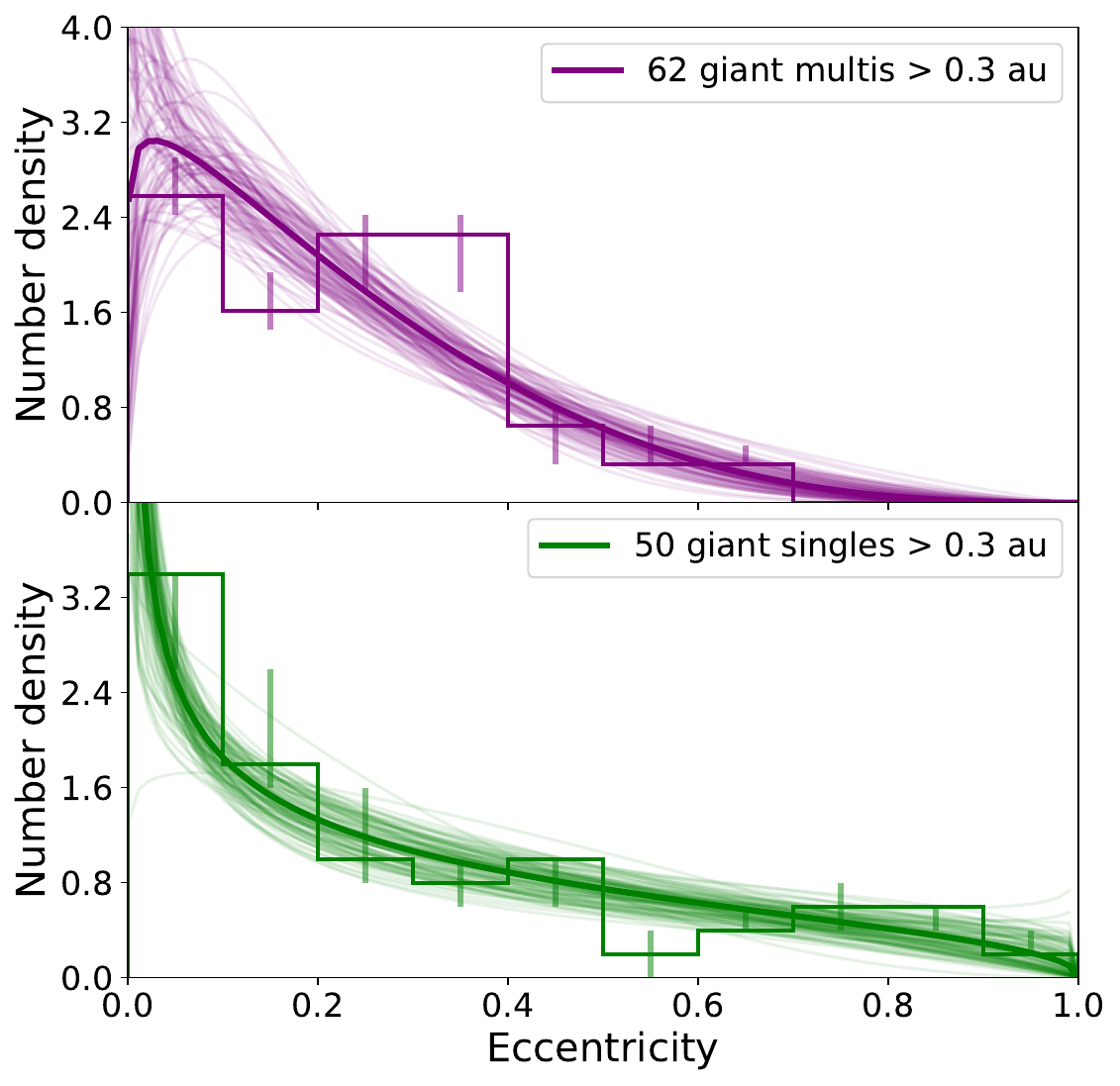}
\caption{Eccentricity distributions of single and multiple giants beyond 0.3 au. Purple/top distribution is for multiple giants, green/bottom distribution is for single giants. Histograms are distributions of individual maximum posterior values. Vertical bars show 68$\%$ confidence intervals}. Curves show the median and many draws from beta distribution posteriors, produced using hierarchical inference.
\label{fig:eccentricity_comparison}
\end{figure}

\begin{figure}[ht!]
\includegraphics[width = 0.5\textwidth]{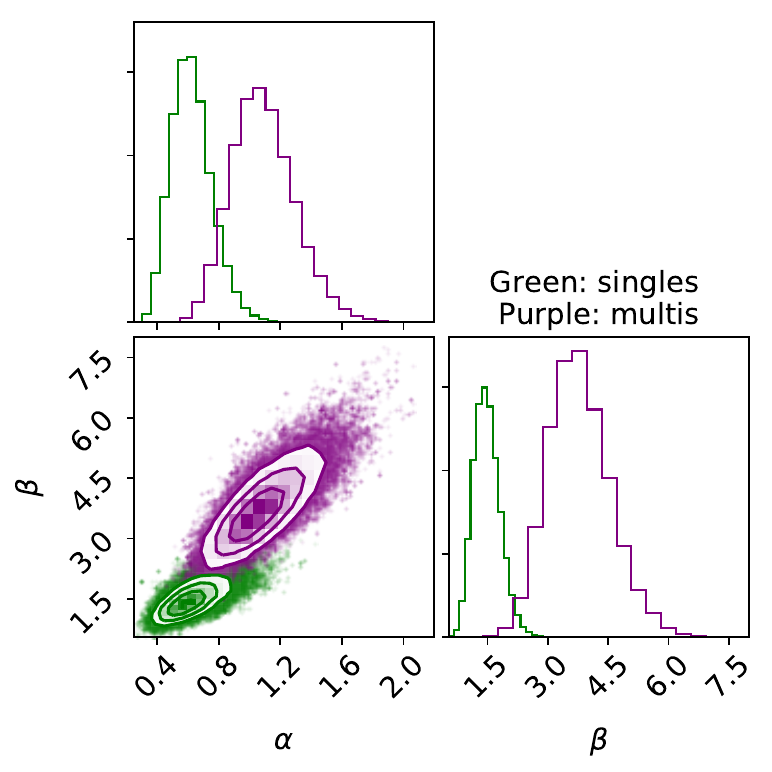}
\caption{Beta distribution posteriors for the single and multiple giant eccentricity distributions. The two populations are $\sim3\sigma$-distinct.}
\label{fig:eccentricity_corner}
\end{figure}

We use the hierarchical model described in Section \ref{sec:methody} to characterize the distinct eccentricity distributions of single and multiple giant planets. Figures \ref{fig:eccentricity_comparison} and \ref{fig:eccentricity_corner} show our results when we fit a beta distribution to these populations. For single giants, our model returns beta parameters $\alpha = 0.60^{+0.14}_{-0.12}$ and $\beta = 1.45^{+0.34}_{-0.30}$. For multiple giants, our model returns $\alpha = 1.06^{+0.21}_{-0.18}$ and $\beta = 3.68^{+0.81}_{-0.70}$. While we define multiplicity as the number of giant planets at any semi-major axis, our joint fits only include planets beyond 0.3 au, since giants within that distance have short tidal circularization timescales \citep{Ogilvie14}. For reference, our population fit to all giant planets beyond 0.3 au returns $\alpha = 0.85^{+0.13}_{-0.11}$ and $\beta = 2.27^{+0.36}_{-0.32}$.

Both the beta distribution models and the raw posteriors show that single giant planets have a pile-up of circular orbits and a long tail that extends to $e \leq 1$, while the multiple giant planets have a more extended range of moderate eccentricity and a sharp cutoff around $e \sim 0.7$. Hints of this behavior are visible in in Figure 13 of \cite{Wright09}, which compared the eccentricity distributions of all single and multi-planet systems discovered to date and found a long single-planet tail not shared by the multis. \cite{Bryan16} reported a similar result with a larger planet sample.

One possible explanation is that multiple giant systems have some probability of experiencing secular interactions or scattering events, whereas truly single giants need strong disk interactions or Kozai-Lidov interactions with a stellar binary to reach high eccentricities. This would imply that the observed population of more circular single giants have not experienced companion-driven excitation, while the moderately eccentric multis and highly eccentric singles represent different outcomes from planet-planet interactions. Eccentric giants that appear to be lonely now may have been born with nearby companions and scattered them into ejection or tidal engulfment, while systems with two observed, moderately eccentric giants experienced dynamical interactions that were strong enough to excite but not eject or cause engulfment. The alternative of Kozai-Lidov excitation by binary stars is entirely possible, particularly since a Gaia query reveals that most of the 12 CLS singles with e $>$ 0.5 have wide stellar companions.

\subsection{Hosts of multiple giants are more metal-rich than single giant hosts}

\begin{figure}[ht!]
\includegraphics[width = 0.5\textwidth]{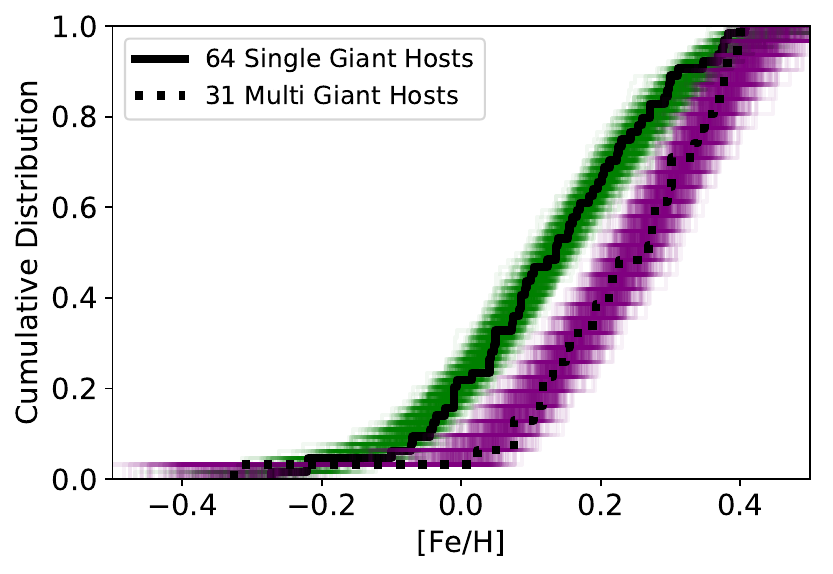}
\caption{Cumulative metallicity distributions for hosts of single giant planets and hosts of multiple giant planets. Solid and dashed steps show distributions of median metallicity measurements, while transparent steps show metallicities drawn many times from Gaussian distributions, with means and standard deviations taken from measurement means and uncertainties. Anderson-Darling tests show that the two groups are distinct.}
\label{fig:metal_single_multi}
\end{figure}

\begin{figure}[ht!]
\includegraphics[width = 0.5\textwidth]{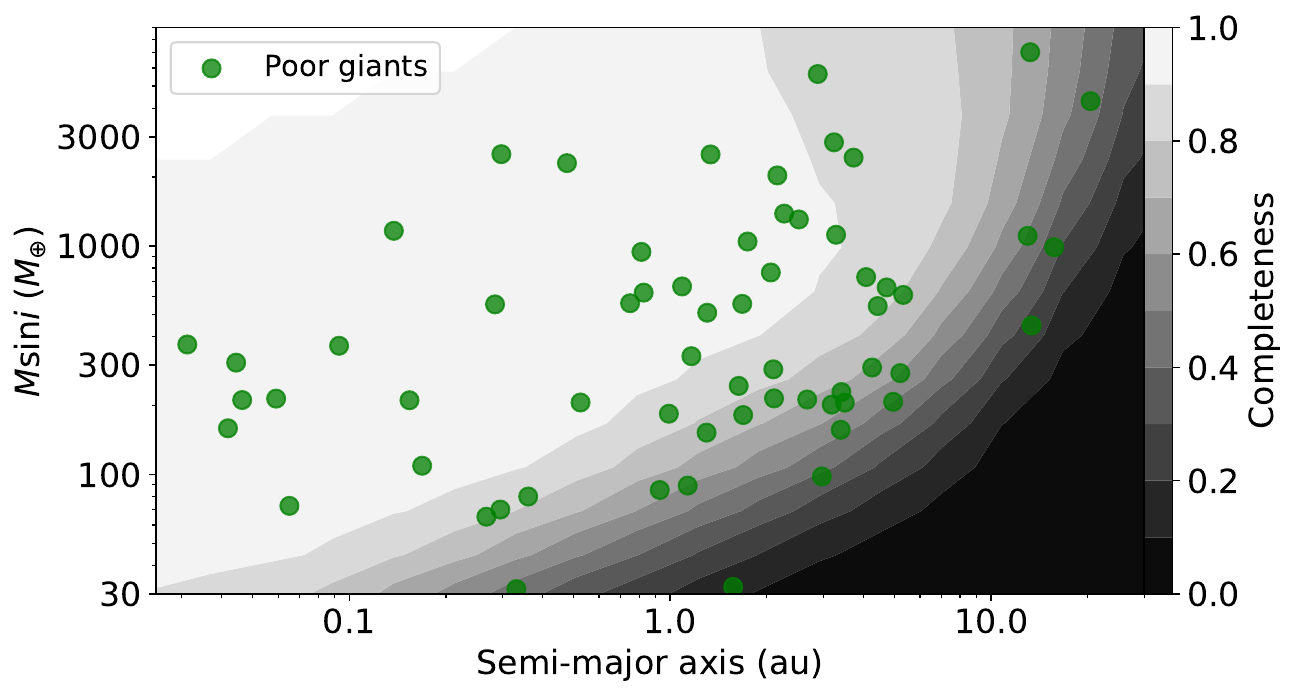} \\
\includegraphics[width = 0.5\textwidth]{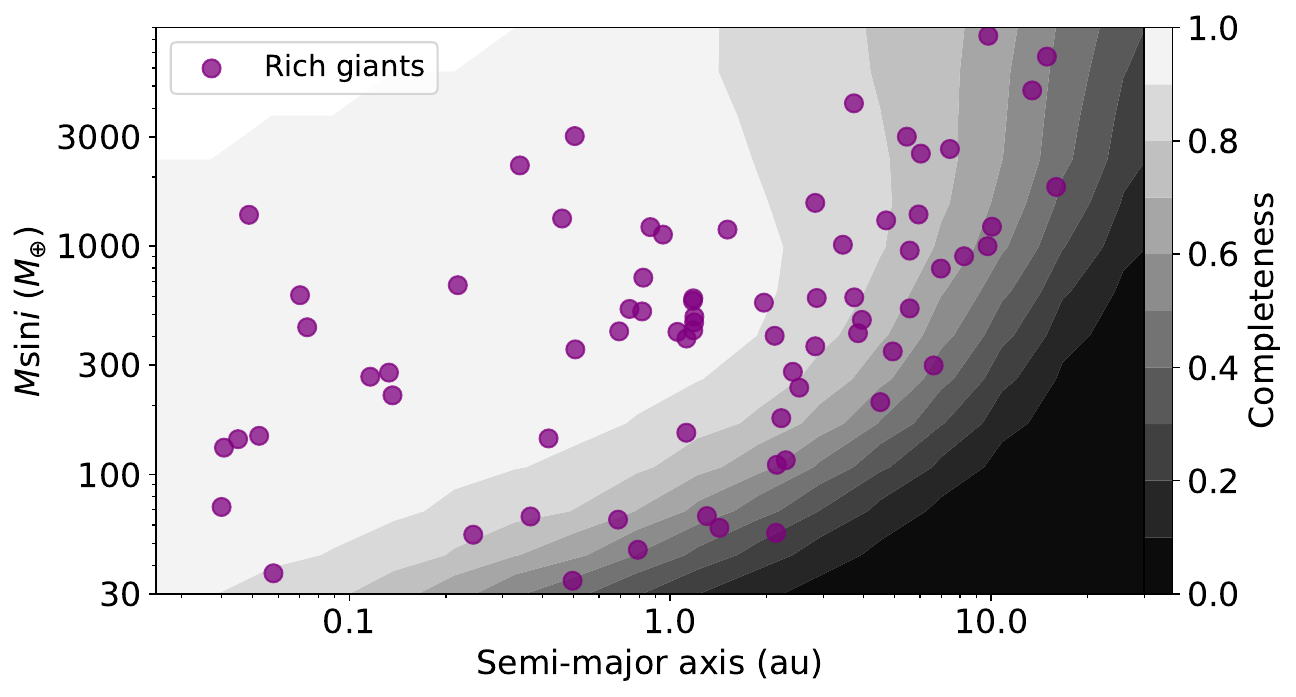}
\caption{Top: Average search completeness contours for giant hosts below their median metallicity. Bottom: Average search completeness contours for giant hosts above their median metallicity. The two contours are nearly equivalent, implying that sensitivity bias to metal-rich stars is not the cause of the single-multi host distinction seen in Figure \ref{fig:metal_single_multi}.}
\label{fig:metal_contours}
\end{figure}

We compare the host star metallicity distributions of systems with only one observed giant planet and multiple observed giant planets to determine whether these two groups are different. Figure \ref{fig:metal_single_multi} shows that hosts of multiple giant planets are distinctly more metal rich than hosts of only one detected giant planet, with 82.2$\%$ significance, that being the fraction of Anderson-Darling tests performed on many samples from our distributions that fall within a p-value $p\leq0.01$. This finding is in agreement with the trend first observed in \cite{Fischer05}. We split our giant host sample along its median metallicity and found that the search completeness contours for the two subsamples are nearly identical, as seen in Figure \ref{fig:metal_contours}. This provides confidence that the difference in these two distributions is not caused by more extensive RV observations of our most metal-rich host stars.

We can interpret this phenomenon in several ways. One possible explanation is that the probability of more than one giant planet forming around a star increases with increasing host metallicity. This is consistent with expectations from the core accretion theory of giant formation \citep{Lissauer09}. Another possible explanation is that metal-richer stars are more likely to form multiple giant planets closer in, and that our sample of single giants contains additional undetected giant companions beyond 30 au. This is disfavored by the results from \cite{Fulton21}, which found evidence that giant planets are less common beyond 8 au than within 1--8 au.

In light of our single-multi metallicity result, we explore whether the eccentricity distribution or semi-major axis distribution of the giant planet population is also metallicity-dependent. The hosts of circular ($e < 0.2$) giant planets and hosts of eccentric ($e \geq 0.2$) giant planets have median metallicities separated by less than 1$\sigma$ in their standard errors, and Anderson-Darling tests produce p-values almost entirely above 0.01. The same is true of hot ($a < 0.1$) giant hosts with respect to colder ($a \geq 0.1$) giant hosts. This rules out the possibility that our single-multi metallicity result is driven by $a$ or $e$ as confounding factors, rather than multiplicity itself.

\begin{figure}[ht!]
\includegraphics[width = 0.5\textwidth]{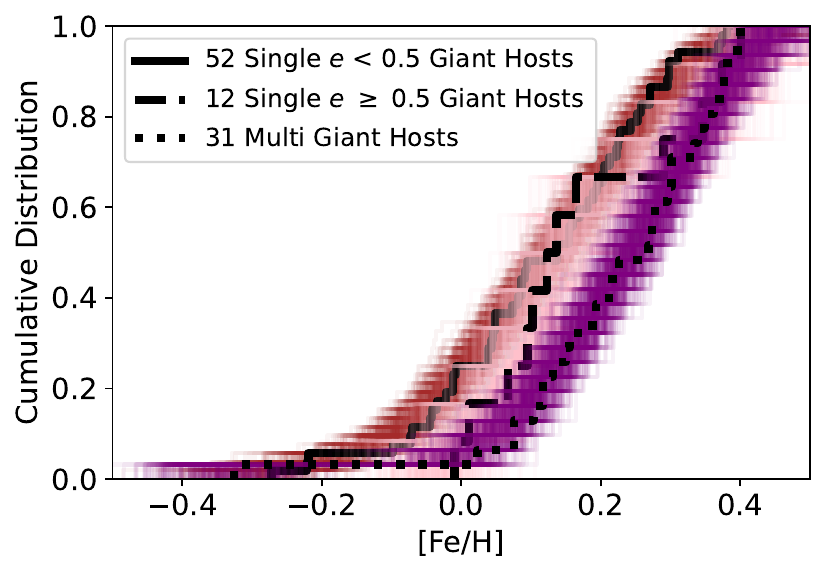}
\caption{Cumulative metallicity distributions for hosts of circular ($e < 0.5$) giant planets, hosts of eccentric ($e \geq 0.5$) giant planets, and multi giant hosts. See Figure \ref{fig:metal_single_multi} for description. The eccentric single hosts are inconclusively drawn from the same underlying metallicity distribution as the less eccentric singles or the multis.}
\label{fig:metal_tail}
\end{figure}

Splitting single giants along $e \geq 0.5$, further along the long tail of eccentric singles, leads to inconclusive results. Figure \ref{fig:metal_tail} implies that there is not a conclusive difference in host metallicity between single $e \geq 0.5$ hosts and hosts of less eccentric singles, and Anderson-Darling tests also show no conclusive difference. The medians of the two groups are separated by just 1.1$\sigma$. However, the same is all true of the difference between single $e \geq 0.5$ hosts and all multi giant hosts; Anderson-Darling tests are inconclusive, and the medians of the two groups are separated by 1$\sigma$. We therefore conclude that our sample of single eccentric planets is not large enough to differentiate its metallicity distribution from that of either the population of single planets on circular orbits or the multi giant population. This is not surprising, as our sample only contains 12 single planets with eccentricities higher than 0.5.  The two comparison samples are proportionally larger, with 52 systems with single giants on low eccentricity orbits ($e$ $\leq$ 0.5) and 31 systems with multiple giant planets.

\subsection{Hot Jupiter pile-up is missing from multis}\label{subsec:warmies}

\begin{figure}[ht!]
\includegraphics[width = 0.5\textwidth]{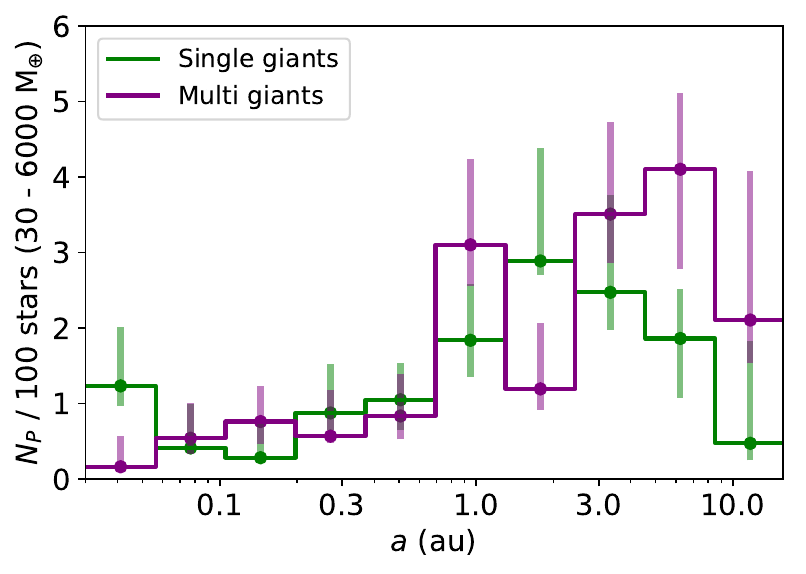}
\caption{Occurrence with respect to semi-major axis for single giant planets and multiple giant planets. Steps and dots are median values, while vertical lines are 15.9--84.1$\%$ confidence intervals. The single giant planets display a pile-up of very hot Jupiters within 0.06 au, tentatively not shared by the multiple giant planets.}
\label{fig:occurrence_axis}
\end{figure}

We use our Poisson likelihood model to compute occurrence as a function of orbital distance for the distinct single giant and multiple giant populations. Figure \ref{fig:occurrence_axis} shows our results. There is a hot Jupiter pile-up within 0.6 au for lonely giants, but no such pile-up for neighborly giants. We further investigate this phenomenon by generating occurrence posteriors for three classes of giants: warm Jupiters (0.1--0.4 au), moderately hot Jupiters (0.06--0.1 au), and very hot Jupiters (0.03--0.06 au). The single giant and multiple giant distributions of warm and hot Jupiters are indistinguishable, but the very hot Jupiter distributions are separated with 97.6$\%$ confidence. We measured this confidence as the fraction of random draws from the difference between distributions that is greater than zero. When we define multiplicity to include all massive companions, including brown dwarfs and stellar binaries, one very hot Jupiter changes in multiplicity status, and our separation confidence decreases to 94.4$\%$ significance.

\begin{figure}[ht!]
\includegraphics[width = 0.5\textwidth]{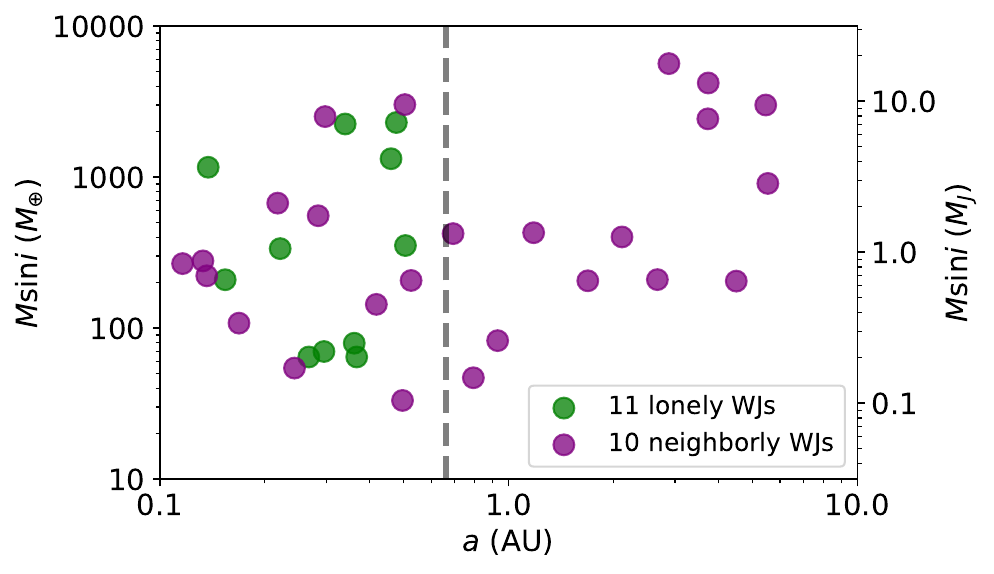}
\caption{Lonely and neighborly warm Jupiters, described in Subsection \ref{subsec:warmies}. 11 systems have lonely warm Jupiters, while 10 systems host both a warm Jupiter and at least one detected outer giant companion. This does not refute the synthetically driven claim by \cite{Jackson21} of secular warm Jupiter migration.}
\label{fig:warmies}
\end{figure}

While very hot Jupiters may be lonely, warm Jupiters ($\geq$ 0.1 au) may tell another story. \cite{Jackson21} predicts that warm Jupiters have outer giant companions with significantly non-zero frequency, which could imply that these companions interact secularly to bring the warm Jupiters to their final states. Alternatively, \cite{Huang16} found that warm Jupiters are more likely to have small planet companions than hot Jupiters. This could imply that warmer giants are less likely to have migrated inward, since migration would disrupt the orbits of small companions. We test these hypotheses with our sample of warm Jupiters, shown in Figure \ref{fig:warmies}. In a sample of 21 systems, 10 show detected outer giant companions, while 11 do not. We satisfy the prediction of \cite{Jackson21}, and therefore cannot rule out secular migration. Further resolving this question requires a sample that is uniformly sensitive to both cold gas giants and small rocky planets, which is not the case for the CLS.

\subsection{Inconclusive mass functions} \label{subsec:mass}
\begin{figure}[ht!]
\includegraphics[width = 0.5\textwidth]{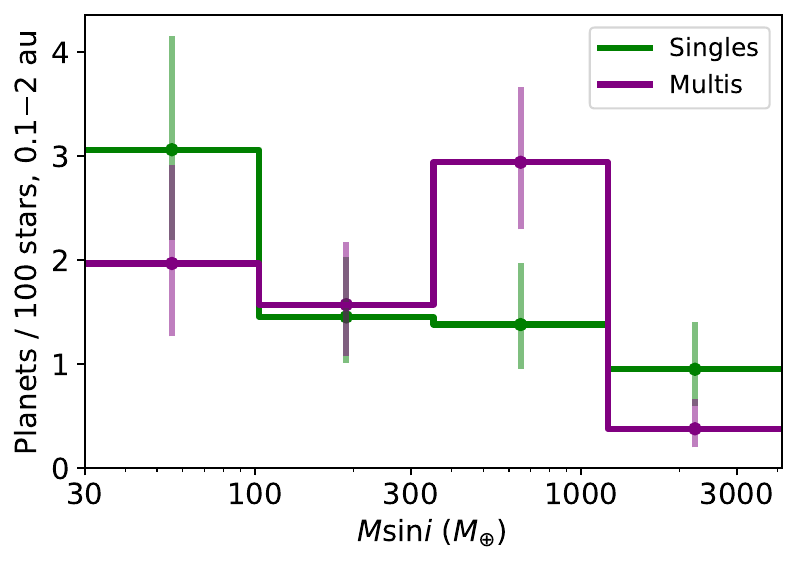}
\caption{Occurrence as a function of \msini\ for single and multiple giants. Steps and dots are median values, while vertical lines are 15.9--84.1$\%$ confidence intervals. The two distributions are not $\leq2\sigma$ separated in all four considered ranges.}
\label{fig:massie}
\end{figure}

By investigating occurrence as a function of mass for single and multiple giant planets separately, we may be able to infer whether the two groups form via different pathways. Figure \ref{fig:massie} shows occurrence rates with respect to \msini\ of both single and multiple giants within 0.1--2 au (multis can include planets outside this range). These two distributions are $\leq2\sigma$ separated in all four considered ranges of \msini\ , but the multi distribution tends toward higher masses than the single distribution. Leaving aside our occurrence model, the median \msini\ of our single giant planets is 0.92 \mjup , while the median \msini\ of giant multiples is 1.71 \mjup . While we cannot assume that the giant masses are normally distributed, we can report that the standard error in the mean \msini\ for singles is 0.33 \mjup , and that the standard error in the mean \msini\ for multis is 0.42 \mjup .

\begin{figure}[ht!]
\includegraphics[width = 0.5\textwidth]{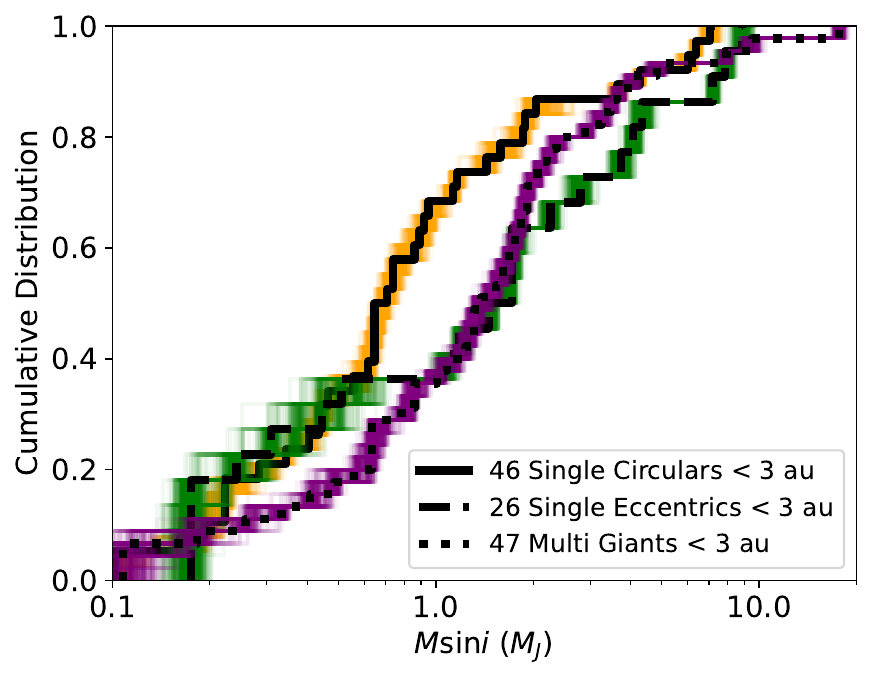}
\caption{Cumulative \msini\ distributions for multiple giants within 3 au and single giants split across an eccentricity of 0.2. While the single and circular giants are $>4\sigma$ separated from the multis, eccentric giants are inconclusively more massive. Note that this may be a selection effect for more eccentric detections with greater mass.}
\label{fig:massie_cumulative}
\end{figure}

Stepping back from our occurrence model, we can compare the cumulative distributions of the two groups and make assumptions that the average completeness contours for the single hosts and multi hosts are nearly identical, similar to the metallicity-split contours in Figure \ref{fig:metal_contours}. Figure \ref{fig:massie_cumulative} shows these cumulative distributions. AD tests show that there is a $>99.7\%$ chance of a p-value below 0.01 for matching single and multi giants within 3 au. This can be interpreted as the degree of confidence in the claim that giants in multiples are consistently more massive than single giants. Additionally, while eccentric singles appear to be more massive than circular singles, this could be a selection effect of more eccentric detections with greater mass.

\subsection{The properties of multiple giant planets}

\begin{figure}[ht!]
\includegraphics[width = 0.5\textwidth]{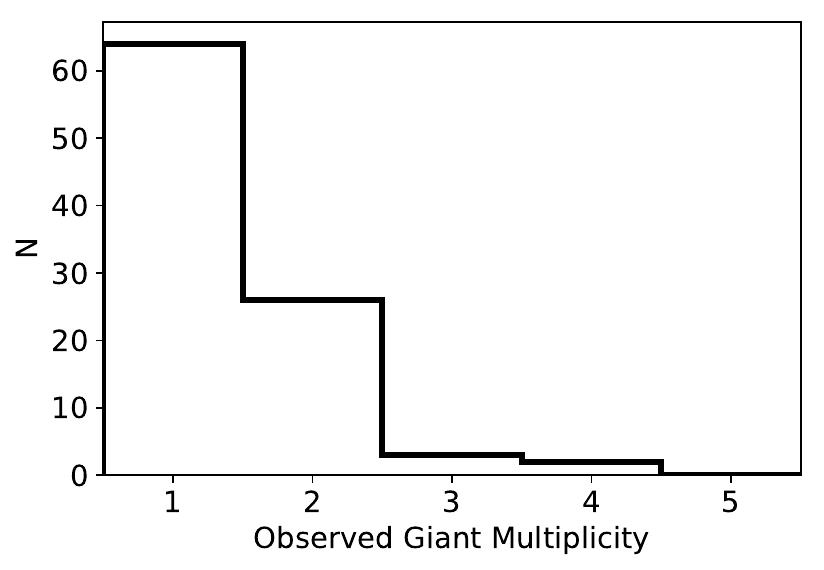}
\caption{Observed multiplicity of giant planets in the CLS sample. Half as many systems contain two or more giants as contain only one giant.}
\label{fig:observed_multiplicity}
\end{figure}

Comparative work requires taking a broader view of giant multiplicity, beyond hot or warm Jupiters. Figure \ref{fig:observed_multiplicity} shows the observed giant multiplicity distribution of the CLS. Approximately two thirds of all stars that host a giant planet host only one observed giant planet, while one third of giant hosts show two or more observed giants. This distribution is not corrected for search completeness; it only shows the multiplicity of detected giant planets. This means that at least one third of stars that host one giant planet also host two or more giant planets, unless we are somehow less sensitive to single giants than to multi giants.

By looking at systems of multiple giant planets, we can also explore how important global disk properties are for setting final planet masses compared to the timeline and location of each individual planet's formation. Focusing on multiple giant systems, we investigate whether paired giant planets have correlated masses. Specifically, we compare $\msini$ values of neighboring giant planets, for each neighboring giant pair in a system. We measure a Pearson correlation coefficient of $R=0.476\pm0.010$ (error come from bootstrapping measured \msini\ values with uncertainties). We use two independent methods to investigate whether sampling a population of uncorrelated giant pairs may produce this value of $R$.

In one case, we use bootstrapping to approximate uncorrelated populations using the observed CLS catalog. First, we repeatedly draw two giant planets from the CLS catalog until we have thirty-one pairs, the same size as the number of observed multi giant systems. We sample without replacement to ensure that we are shuffling our entire giant catalog, rather than drawing the same planets multiple times. Then, we measure the Pearson correlation coefficient of this shuffled population. We also measure the standard deviation in the log-ratio of planet pair $\msini$s, and assume coplanarity to simplify this to $\sigma_{\mathrm{ln}(M_2/M_1)}$. It must be noted that coplanarity between giant planets is not a given, and that this limits the significance of our results.

We repeat this process 10$^5$ times, until we build up bootstrapped distributions of $R$ and $\sigma_{\mathrm{ln}(M_2/M_1)}$. Our measured $R$ falls just outside the 99.7$\%$ confidence interval of the bootstrapped distribution, at the 99.71$\%$ confidence interval. This means that there is a 0.29$\%$ probability that an uncorrelated giant planet population could produce a Pearson correlation coefficient of 0.45. Our measured $\sigma_{\mathrm{ln}(M_2/M_1)}$ falls on the 1.4$\%$ confidence interval of the bootstrapped distribution.

We also use a synthetic population experiment to independently measure the significance of our observed correlation. We drew giant planet pairs from a uniform ln(\msini\ ) distribution and the semi-major axis distribution reported in \cite{Fulton21}, with a minimum period semi-major axis ratio ln($a_2/a_1$) $>$ 0.5. We impose this limit on the semi-major axis ratio in order to prevent unlikely or unphysical draws of pairs within 3:2 resonance. We took our search completeness contours into account to only select simulated planet pairs for which both planets are detected. We detail our synthetic population algorithm in the list below.

\begin{itemize}
  \item Draw two values of ln(\msini\ ) from a uniform distribution between ln(30 \mearth)--ln(6000 \mearth), and two values of ln($a$) from the nonparametric giant planet model reported in \cite{Fulton21}, bounded between 0.1--10 au. We approximate this distribution as piecewise-uniform in ln($a$), with a break at 1 au, integrated probability of 2/11 between 0.1--1 au, and integrated probability of 9/11 between 1--10 au.
  \item Measure the search completeness and draw a random number from a uniform 0--1 distribution for each set of ln($a$) and ln($\msini$). For each set, if the random number drawn is less than the detection probability at those parameters, report the planet corresponding to that set as detected.
  \item Keep the planet pair if we detect both planets and ln($a_2/a_1$) $>$ 0.5, otherwise reject it.
  \item Repeat until we have drawn the desired number of simulated planet pairs for one synthetic population.
  \item Repeat until we have drawn the desired number of synthetic populations.
\end{itemize}

With $10^4$ synthetic populations of 31 observed planet pairs each, this test finds that our observed correlation falls just within the 99$\%$ confidence interval of the distribution that our uncorrelated simulations produces.

\section{Discussion}

\subsection{Metallicity, eccentricity, and multiplicity}

\cite{Dawson13} found that giants with metal-rich hosts show signatures of planet-planet scattering. In particular, giant planets between 0.1 AU and 1 AU are likely to be more eccentric when they orbit stars with [Fe/H] $>$ 0. This higher eccentricity could be caused by planet-planet scattering events, which would imply that metal-rich stars are more likely to host multiple giant planets. Our results support this hypothesis, as we find that systems of multiple giant planets are more common around metal-rich stars, and that systems with multiple giant planets have higher average eccentricities than those in single giant planet systems. Table \ref{tab:comparisons} shows the above properties side-by-side for the two populations.

\begin{longtable*}{lrrr}
\toprule
\midrule
Property & Characteristic &  Single Giant Planets & Multiple Giant Planets \\
\toprule
\textbf{Eccentricity}\\
& 10th percentile, $e_{10\%}$ & 0.03 & 0.03 \\
& Median, $e_{50\%}$ & 0.20 & 0.23  \\
& 90th percentile, $e_{90\%}$ & 0.77 & 0.47 \\
& Beta distribution $\alpha$ & $0.60^{+0.14}_{-0.12}$ & $1.06^{+0.21}_{-0.18}$  \\
& Beta distribution $\beta$ & $1.45^{+0.34}_{-0.30}$ & $3.68^{+0.81}_{-0.70}$ \\
\textbf{Semi-major Axis} \\
 & Comparison & HJ pile-up & HJ deficit  \\
\textbf{Mass} \\
& Median & 0.92 \mjup & 1.71 \mjup \\
& Mean & 2.07 $\pm$ 0.33 \mjup & 2.88 $\pm$ 0.42 \mjup \\
& Feature & Less massive & Intra-system uniformity \\
\textbf{Iron Abundance} \\
& $<[$Fe/H$]>$ & $0.129 \pm 0.019$ dex & $0.228 \pm 0.027$ dex\\
& Comparison & Enriched over Solar & More enriched \\
\textbf{Stellar Binarity} \\
& Fraction Bound (Lit.) & $26.2\pm6.3\%$ & $25.8\pm9.1\%$ \\
\bottomrule
\caption{Summary of population comparison between single giant systems and multi giant systems. Fraction of bound binaries comes from a literature estimate with Simbad and the Washington Double-star Catalog. Uncertainties come from Poisson counting error.\label{tab:comparisons}}
\end{longtable*}

Although this story is clear, our data are less definitive on the question of how the sub-population of single giants on highly eccentric orbits acquired their eccentricities. There are few other mechanisms that could explain the metal-poor, high-eccentricity tail of lonely giants above $e \sim 0.6$. The most likely alternative to scattering is the Kozai-Lidov mechanism, by which a massive outer companion secularly perturbs the orbit of an inner companion, can potentially excite giants to $e > 0.6$, but only given a high mutual inclination and absence of other dynamical factors \cite{Naoz16}. Even if this could account for all of our single, highly eccentric giants, it would require the presence of stellar binary companions or undetected giant companions in these systems. We went back and examined the radial velocity data sets for the eccentric single planets in our sample, and found that 4 out of 65 single giant systems have $\geq3\sigma$-significant parabolic or linear trends, and two of these (HD 34445 and HD 156668) show strong linear correlation between their RV trends and S-values, which we can treat as a proxy for magnetic activity. Therefore, only 2 out of 65 single giant systems show RV evidence for a stellar binary companion (HD 195019 and HD 145934), and neither of these systems host giants with $e\geq0.2$. HD 195019 has a visually resolved wide-orbit binary companion \citep{Lu87}, while HD 145934's status is unclear.

\begin{figure}[ht!]
\includegraphics[width = 0.5\textwidth]{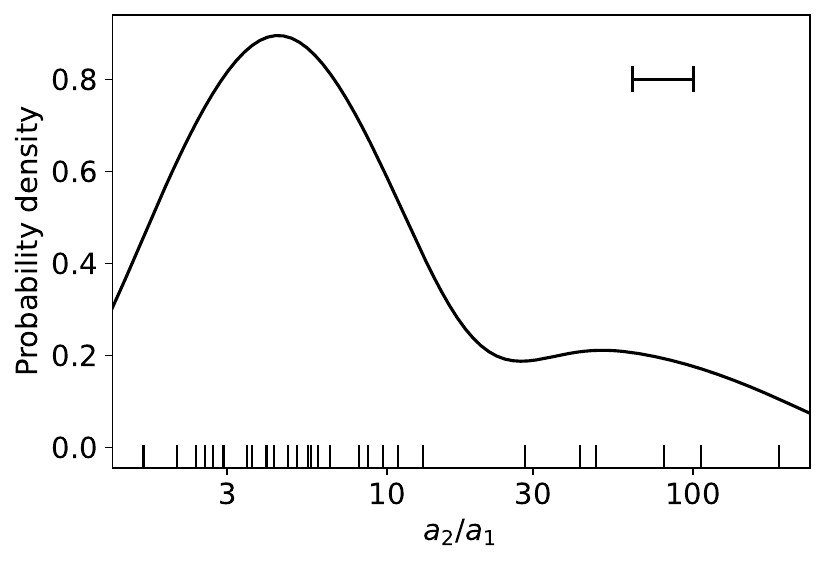}
\caption{Observed semi-major axis ratio distribution of two-giant systems. The curve is a kernel density estimation of the distribution, while the vertical ticks show the observed set of axis ratios. Horizontal bar shows the kernel width.}
\label{fig:axis_ratios}
\end{figure}

Alternatively, it is possible that most of the single planets in our sample do in fact have additional sub-Jovian giant companions located outside 10 au. Our survey may be missing giant planets beyond this distance, and therefore misidentifying multi giant systems as single giant systems. As seen in Figure \ref{fig:axis_ratios}, the  majority of CLS-detected giant planet pairs have a semi-major axis ratio less than 10, with a peak ratio frequency around 3. This means that our coldest single giants around 10--20 au could have companions out around 30--60 au, where they would be undetectable if they were small enough and on circular orbits. Again, this is disfavored by the giant planet fall-off beyond 8 au reported in \cite{Fulton21}. Additionally, there could be highly-inclined giant planets in our sample that evade RV detection and enhance the apparent population of highly eccentric singles.

Where do our lonely and highly eccentric giants come from? Out of 50 CLS single giant planets beyond 0.3 au, 11 have median eccentricities greater than 0.5. CLS contains 30 multi giant systems with at least one giant beyond 0.3 au. If the eccentric singles all used to be scattered multis, that would imply that 11 out of 41 multi systems, or roughly a quarter, scattered themselves into singledom. \cite{Carrera19} found that the observed eccentricity distribution of giants within 5 au is compatible with scattering accounting for all giants with $e\geq0.3$. This supports the claim that many highly eccentric singles used to be multis that experienced scattering events. However, our poorly constrained metallicity distribution for high-eccentricity giant hosts limits our ability to make claims about their original multiplicity.

We leave this discussion with more questions than when we began, but are hopeful that future work will elucidate these mysteries. Longer RV baselines for eccentric single systems and imaging campaigns to look for distant stellar or substellar companions could clarify the statuses of our single and eccentric giants, while synthetic population models might have something interesting to say about the origins of our observed eccentricity and metallicity distributions.

\subsection{Ultra-hot Jupiter pile-up}

\cite{Knutson14} performed an RV-trend search for massive companions to hot Jupiters, and found that $50\pm10\%$ of giants within 0.7--11 days host an outer companion between 1--13 \mjup\ and 1--20 au. More than two-thirds of their hot Jupiters with companions have orbital periods within 6 days. In the CLS sample, 3 out of 15 hot Jupiters within 10 days host an outer companion in the same range (HD 187123, HD 217107, HD 9826). That gives a lower bound on the companion rate of $20\pm12\%$.

These two rates are compatible within $2.14\sigma$, and therefore consistent with the companion occurrence rate reported in \cite{Knutson14}. In order to make a more direct comparison we would need to calculate the completeness-corrected companion rate for the hot Jupiters in the Legacy sample, but this is outside the scope of the current study.  It is also worth noting that many of the companions reported in \cite{Knutson14} were detected as linear trends with 5-8 year baselines.  These systems would benefit from additional RV follow-up in order to more rigorously explore potential false positive scenarios due to stellar activity and better constrain the masses and orbital semi-major axes of the companions.

\subsection{Intra-system mass uniformity}
Previous studies of sub-Neptune-sized transiting planets have found that planets located in the same system have correlated masses and radii \citep{Weiss18, Millholland17}.  Our study reveals that giant planets forming in the same system also appear to have correlated masses. This could imply that the global properties of protoplanetary disks strongly influence the final masses of their nascent giant planets, and that these properties are more important than solid surface density and other properties that vary radially or over time. That condition could be initial protoplanetary disk mass, since disk mass determines how much material is available during giant planet formation, solid mass for large core formation, or disk lifetime, since quickly dissipating disks have less time to make giants.

\cite{Weiss18} measured a Pearson correlation of $R=0.65$ for the paired planet radii of a larger sample than the CLS. This means that the magnitude of the `small peas in a pod' effect is significantly greater than that of the `giant peas in a pod' effect. The final masses of giant planets are very sensitive to the relative timing of the onset of runaway accretion. Giant planets can also cut of the flow of gas to the inner disk, which could affect the ability of an inner companion to accrete a massive gas envelope. This might explain the weaker correlation between giant masses.

\section{Conclusions}

We compared and contrasted single giant systems and multi giant systems, and found several key differences. When we look at giant occurrence with respect to orbital separation, giant singles have a super-hot Jupiter pile-up not shared by giant multis, limited by lack of sensitivity to giant planets beyond 20 au. We reproduced previously seen eccentricity distributions for these two groups, and found that the single giants have a circular orbit pile-up and long eccentric tail, while the multis are spread across moderate eccentricities and limited to $e \leq 0.7$. We also found that multi giant stellar hosts are more metal-rich than single giant hosts, and that total giant planet mass (inferred from \msini\ ) does not correlate with host metallicity. Taken together, these results may imply that more metal-rich stars form more discrete giant planets, possibly due to formation of more rocky cores. We also found that at least a third of stars that host giant planets host two or more of them. This implies that giant planet pairs probably have correlated masses, perhaps that the shared formation environment of giants influences their final mass limits.

Although this work provides new information about the underlying population of giants, it also raises unresolved questions about why we see these patterns. What does the single super-hot Jupiter pile-up mean for theories of hot Jupiter formation? Are lonely and highly eccentric giants products of planet-planet scattering? And why are giant planet pair masses correlated? These questions can be the starting points for new theoretical inquiries into planet formation.

\acknowledgments
L.J.R.\ led the construction of this paper, including performing all analysis, generating all of the figures, and writing this manuscript. H.A.K. and A.W.H. advised substantially on the scientific direction of this work. B.J.F. collaborated on occurrence analysis.

A.W.H.\ acknowledges NSF grant 1753582. H.A.K. acknowledges NSF grant 1555095.

We thank Fei Dai, Lauren Weiss, and Konstantin Batygin for valuable discussions.

This work has made use of data from the European Space Agency (ESA) mission {\it Gaia} (\url{https://www.cosmos.esa.int/gaia}), processed by the {\it Gaia} Data Processing and Analysis Consortium (DPAC, \url{https://www.cosmos.esa.int/web/gaia/dpac/consortium}). Funding for the DPAC has been provided by national institutions, in particular the institutions participating in the {\it Gaia} Multilateral Agreement.

\vspace{5mm}

\software{All code used in this paper is available at \url{github.com/California-Planet-Search/rvsearch} and \url{github.com/leerosenthalj/CLSIV}. This research makes use of GNU Parallel \citep{Tange11}. We made use of the following publicly available Python modules: \texttt{astropy} \citep{Astropy-Collaboration13}, \texttt{matplotlib} \citep{Hunter07}, \texttt{numpy/scipy} \citep{numpy/scipy}, \texttt{pandas} \citep{pandas}, \texttt{emcee} \citep{DFM13}, and \texttt{RadVel} \citep{Fulton18}.}

\bibliographystyle{aasjournal}
\bibliography{CLSIV}{}

\end{document}